# Mutual Information based Bayesian Analysis of Power System Reliability

Swasti R. Khuntia, *Graduate Student Member, IEEE*, José L. Rueda, *Senior Member, IEEE* and
Mart A. M. M. van der Meijden, *Member, IEEE*

*Abstract*—This paper aims at assessing the power system reliability by estimating loss of load (LOL) index using mutual information based Bayesian approach. Reliability analysis is a key component in the design, analysis and tuning of complex structure like electrical power system. Consideration is given to rare events while constructing the Bayesian network, which provides reliable estimates of probability distribution function of LOL with lesser computing effort. Also, the ranking of load components due to loss of load is evaluated. The RBTS and IEEE RTS-24 systems are used as test cases.

## Nomenclature

| | |
|---|---|
| LOL | Loss of load |
| M | Mathematical formulation of Bayesian network |
| S | Structure of Bayesian network |
| θ | Parameters of Bayesian network |
| $C_i$ | Load curtailment vector |
| $W_i$ | Weighting factor for load |
| NC | Set of load buses |
| $S^j$ | System state in $j$th state |
| $T(S^j)$ | Line flow vectors under state $S^j$ |
| $A(S^j)$ | Relation matrix between line flows and power injections under state $S^j$ |
| PG | Generation output vector |
| C | Load curtailment vector |
| PD | Load power vector |

## I. Introduction

RELIABILITY assessment of electrical power system is not something very novel, but it still lacks a coherent probabilistic treatment of uncertain data and parameter estimates during contingency. The primary objective of power system reliability assessment is improving reliability by identifying weak points in a network in order to provide qualitative analysis and various reliability indices [1]. Reliability assessment methods have appeared many decades ago starting from generation reliability, and then for transmission reliability [2].

Uncertainties in modern power systems such as load fluctuation, renewable energy sources, faults on transmission lines, etc. increase the importance of probabilistic methods in reliability assessment. Nowadays, power systems are increasingly very large and comprise of a huge number of components. And, accordingly, the uncertainty related to it increases. Despite the efficiency of Monte-Carlo (MC) simulation, evaluation of their reliability using MC can take considerably long time even for the moderate level of precision. State enumeration method would also entail computational bottlenecks due to the very high number of states with increasing number of input variables and associated probabilities. Analytical probabilistic approaches, such as cumulate methods, Gram-Charlier series, point estimation methods, and probabilistic collocation method entail a reduced computing effort, but at expense of imprecision, i.e., misleading estimate of statistical properties of output variables.

Literature survey shows that machine learning is a promising field in electrical power system, which can be applied in assessing the power system reliability [3]-[5]. Two different but efficient machine learning classification methods are artificial neural networks (ANN) and Bayesian networks. ANNs have the disadvantage of not having symbolic reasoning and semantic representation. An ANN generally takes the shape of a ''black box'' model in the sense that the non- linear relationships of cause and effect are not easily interpretable, making it difficult to explain the results. On the other hand, the main advantage of Bayesian network is that reasoning is based on a real-world model. The system has a thorough understanding of the processes involved, rather than just a mere association of data and assumptions. This is combined with a strong probabilistic theory enabling Bayesian approach to give an objective interpretation. Application of ANN in power system reliability studies is featured in various literatures [6]-[8]. Similarly, Bayesian networks have found wide range of applications in power system such as, fault diagnosis [9]-[11], reliability assessment [12]-[15], outage management [16]. This paper aims at evaluating power system reliability based on Bayesian approach with the help of MC-simulation, where the latter is responsible for data generation using mutual information technique.

The rest of this paper is organized as follows. Section II introduces the concept of Bayesian networks, and its

The research leading to these results has received funding from the European Union Seventh Framework Programme (FP7/2007-2013) under grant agreement No 608540 GARPUR project http://www.garpur-project.eu

Swasti R. Khuntia and José L. Rueda are with Intelligent Electrical Power Grids Research Group of Electrical Sustainable Energy Department at Delft University of Technology (TU Delft), 2600GA Delft, The Netherlands (e-mail: s.r.khuntia@tudelft.nl).

Mart A. M. M. van der Meijden is with Intelligent Electrical Power Grids Research Group of Electrical Sustainable Energy Department at Delft University of Technology (TU Delft), 2600GA Delft, and also with TenneT TSO B. V., Arnhem, The Netherlands application and advantages in electrical power system. Section III presents the algorithm and steps for Bayesian analysis while introducing other important terms like importance sampling, and mutual information. The study and evaluation of Bayesian application is concluded in Section IV.

## II. BAYESIAN ANALYSIS

The Bayesian belief net or simply, Bayesian network is a probabilistic graphical model in which a problem is structured as a set of variables or parameters and probabilistic relationships among them [17]. It consists of two parts, namely, structure and parameters. Mathematically, it can be written as $M = (S, \theta)$, where $S$ refers to the structure and $\theta$ refers to the parameters or conditional probability distributions of the Bayesian network. The structure is a Directed Acyclic Graph (DAG) where the nodes represent variables of interest and the links between them indicate information or causal dependencies among the variables. Each of the variables in the network has a finite set of mutually exclusive states, like failure or operating states. Parameters refer to Conditional Probability Distributions (CPD) assigned to the nodes that define probabilistic relationship between each node and its parents [18].

As Jensen [17] states, the structure and parameters of Bayesian network are such that it defines a unique joint probability distribution over variables and hence it avoids the need for a joint probability distribution table of variables whose size increases exponentially when the number of variables increases. The Bayes' rule which forms the basis of Bayesian analysis is given as

$$P(A\mid B)=\frac{P(B\mid A)P(A)}{P(B)} \quad (1)$$

which states the method for updating the beliefs about an event $A$ given that information about another event $B$ is provided. In this formulation, $P(A)$ refers to prior probability of $A$, $P(A|B)$ refers to posterior probability of $A$ given $B$ and $P(B|A)$ refers to likelihood of $A$ given $B$.

Based on above formulation, the Bayesian network considered in our study is shown in Fig. 1. The data-set consists of state vectors as nodes represented as Generator (G), Line (L), Bus (B) and LOL. Data generation is explicitly described in sub-section B of section III.

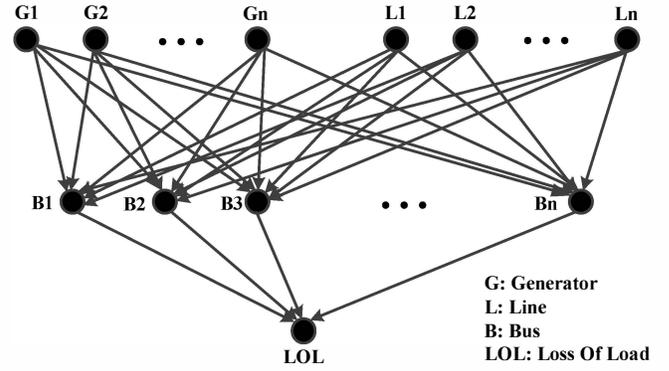

Fig. 1. Schema of considered Bayesian structure

## III. RELIABILITY ASSESSMENT USING BAYESIAN ANALYSIS

To assess the robustness of mutual information based Bayesian approach, and portray a clear understanding of the methods involved, a flowchart is shown in Fig. 2. The first step is generating the data base of numerical simulations by random sampling and to derive system state information for each simulation. In the next step, importance sampling technique is used to extract the knowledge from the generated data. Then, mutual information is employed to assess the linkage in the data set, and subsequently used in decision making and interpreting the relation among different components of the network.

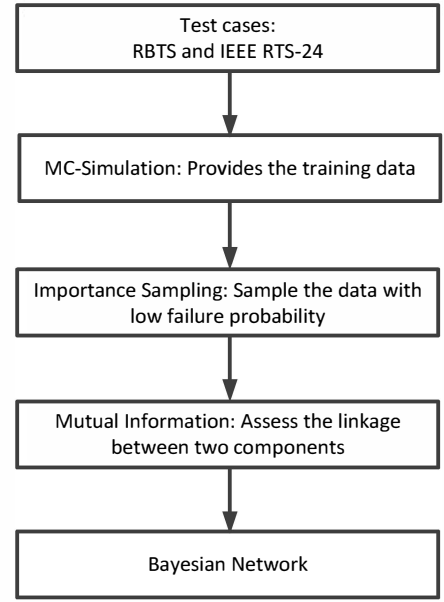

Fig. 2. Step-by-step guideline for constructing the Bayesian network

### A. Test Cases under Study

For assessing the power system reliability, RBTS and IEEE RTS bus systems were chosen. First test case is the Roy Billinton Test System (RBTS) [19] shown in Fig. 3, and later the test was validated for the IEEE RTS-24 bus system [20]. Monte-Carlo (MC) simulation, explained in later sub-section, was used to generate training data for reliability analysis.

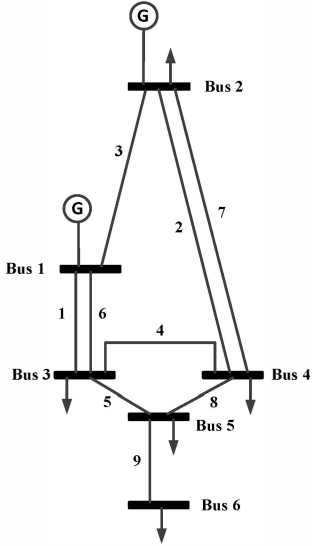

Fig. 3. First test case: Single line diagram of RBTS

*B. MC-Simulation based Data Generation*

The training data is generated by Monte-Carlo (MC) simulation [21]. MC-simulation is chosen because of its wide acceptance for reliability studies and also because of the size of the studied systems. MC is generally classified into two techniques—sequential and non-sequential (random sampling) simulations. Sequential MC simulates the artificial chronological history of all components in correspondence with their probability distributions. This technique operates in the time domain and is therefore capable of directly evaluating every type of indices as well as automatically incorporating correlation between components if any. Sequential MC is a very flexible technique; however, it is known to be computationally expensive. Non-sequential simulation technique, in contrast with sequential MC, neglects chronological histories of components. It randomly samples system state according to its probability of occurrence. This technique, in general, reaches convergence faster than sequential MC though it needs additional computation when calculating frequency and duration (F & D) indices. In our study, components considered are generators, transmission lines and buses as stated in section II. They have only two states of operation, i.e., up/normal or down/failure. The data-set structure is

$$[G1\ G2 \ldots Gn\ L1\ L2 \ldots Ln\ B1\ B2 \ldots Bn\ LOL] \quad (2)$$

LOL refers to loss of load event, which is considered one unless there is zero load supply. The dataset is refined for more factual analysis of components, which have very low failure probabilities using Importance sampling, which is explained later. For the training data, preventive and corrective actions are considered while deciding the state. During preventive action, random number between {0,1} is generated by generating unit, which is compared with the forced outage rate (FOR). If the random number is greater than FOR, it's an up or normal state, equivalent to zero. Else, it is a down or failure state, equivalent to one. During corrective actions, generation rescheduling and load shedding as described in [21] is followed. And, the optimization problem is modified by adding a weighting factor to the load curtailment vector. Thus, the new optimization problem is

$$\min \sum_{i \in NC} W_i C_i$$

such that

$$T(S^j) = A(S^j)(PG + C - D)$$
$$\sum_{i \in NG} PG_i + \sum_{i \in NC} C_i = \sum_{i \in NC} PD_i$$
$$PG^{\min} \leq PG \leq PG^{\max}$$
$$0 \leq C \leq PD$$
$$|T(S^j)| \leq T^{\max}$$

In the above formulation, load is considered constant, and convergence of LOL is chosen as stopping rule in sampling process.

*C. Importance Sampling*

Importance sampling [22] is employed in our study for accurate analysis of transmission system when there is a low probability of failure rate of components. It has been successfully tested in security assessment [21], reliability studies [22] and risk assessment for cascading failures [23]. The importance sampling technique is useful because the independent events with greater effect on results can be identified by changing the associated probability density function. In such case, unlikely events become more likely and reliability assessment provides better results. As explained in previous section, preventive and corrective actions are considered while deciding the component state for data generation.

*D. Mutual Information-based Bayesian Analysis*

The final step in Bayesian analysis is creating the Bayesian network. After considering the rare events through importance sampling, mutual information [24] is used to ascertain the stronger dependencies between nodes. It is a simple and natural measure of dependency. To determine the stronger dependencies between nodes and to eliminate the edges corresponding to weaker dependencies, mutual information technique is employed. Mutual information (MI) is formulated as

$$MI(X,Y) = \sum_{X,Y} P(X,Y) \log\left(\frac{P(X,Y)}{P(X)P(Y)}\right) \quad (3)$$

where, $X$ and $Y$ are discrete random variables, and $P$ refers to observed frequency of data-set samples. MI between two discrete variables is always non-negative and zero when both are independent. Also, there can be a parent-child relationship when there is a strong dependency between $X$ and $Y$.

In our study, mutual information is first calculated between the load point nodes, and LOL node to leave out less important load buses. The parameters for load point nodes are calculated by applying the concept of maximum likelihood [17] to the generated data. Then it is applied for other components, and, it is repeated for both test cases. This leads to the Bayesian network structure, shown in Fig. 4 and Fig. 7.

## IV. SIMULATION RESULTS

The Bayesian approach was validated for RBTS and IEEE RTS-24 bus system using MATLAB [25] on a system with the following configuration: Intel(R) Xeon(R) 8GB 3.7GHz. The model implementation is performed using the BNT toolkit [26], and its toolbox extension for MATLAB is available online [27]. The advantage of using BNT toolbox is the user friendly environment and easy adaptation with MATLAB, as compared to other packages like GeNIe & SMILE developed at University of Pittsburgh [28]. In the BNT platform, the Bayesian network is constructed using a graph structure and corresponding parameters, i.e., conditional probability distribution between nodes and its parameters. The structure of data-set used in this study is shown in Eq. 2. A step-by-step method for constructing the structure is shown in Fig. 2. For both the test systems, in the first step, MC-simulation is performed on the system to obtain the training data. Then the same training data is used to construct the Bayesian structure, shown in Fig. 4 and Fig. 6 respectively. And, this action is executed only once. Thereafter the network may be used for different inferences many times. Evaluation of each of the test systems is explained individually.

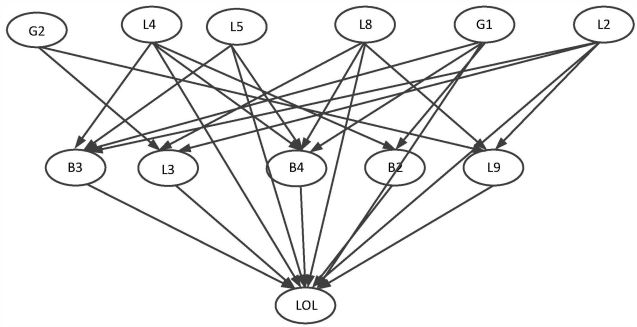

Fig. 4. Bayesian structure of RBTS system

In the first case, Bayesian approach is evaluated on the Roy Billinton Test System (RBTS), which consists of 6 buses that includes 4 load buses, 11 generating units, 9 transmission lines, and 5 load points. A single line diagram is shown in Fig. 3. The RBTS was developed by the Power Systems Research Group at the University of Saskatchewan as a tool for reliability education [19]. The corresponding Bayesian structure shown in Fig. 4 is constructed in 0.82s. The system is analyzed in full load condition. In the figure, lines 4, 5, 8 and generator 1 are parents of bus 4, which is in regard to the load curtailment policy. The policy states that the load curtailment takes place at buses that are close to the elements featured as possible outage component. It also features another characteristic of Bayesian network, i.e., inverse analysis of events. The use of Bayesian network makes it possible to achieve the causes from the effect and hence, identifying the background of events. The most probable components outage event given the loss of load in bus 4 during full load condition is shown in Fig. 5. It is evident that lower reliability of lines and also less generating reserve units induce the outage of generating units. The results might be predictable from the topological characteristics of the system since it is a small system. But, it gets difficult and complicated when extended to a larger system as explained for IEEE-RTS later.

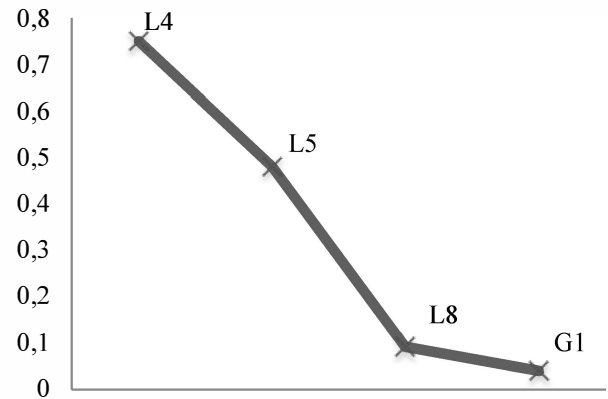

Fig. 5. Component ranking for bus 4 at full load condition

The RBTS consists of 4 load buses. When the cause for loss of load is transmission lines, it is important to identify the load buses with larger fault probabilities. There are two approaches: one is calculating the fault probability of each load bus ($P(B=F)$) and the other is calculating the load probability of one bus when the loss at another load bus is known ($P(B=F|LOL=T)$). The later is preferred due to Bayesian approach because of error in bus-ranking process with the first approach. Table I illustrates the ranking of load buses of RBTS.

TABLE I RANKING OF LOAD BUSES OF RBTS

| Bus | 3 | 5 | 6 | 4 |
|---|---|---|---|---|
| ($P(B=F|LOL=T)$) | 0.65 | 0.53 | 0.30 | 0.22 |

Evaluation of Bayesian approach was performed for IEEE RTS-24 bus system, which consists of 24 buses that includes 17 load buses, 32 generating units, 33 transmission lines, and 5 transformers. When the system is analyzed for peak load condition, the resultant Bayesian structure as shown in Fig. 7 is constructed in 0.95s. It is observed that the Bayesian network consists largely of generation components as compared to transmission components. This is due to the differences in FOR values for generation and transmission components, when fault probability of generation components is dominating. Also, closer look at the structure reveals that it does not include all components and load points, but only critical components as computed by mutual information. Thus, a downside of this approach is that in a highly reliable system, the structure would be complex with a large number of components. As a consequence, it is observed from the structure that due to the presence of local generation or multiple transmission components; few load buses are missing.

The component ranking test was repeated for this test case, and the result for bus 13 is shown in Fig. 8. Thus the loss of load in full load conditions is due to the lower reliability of generating units and also less generating reserve units which induce the outage. It is in contrast to the evaluation for RBTS where lines dominate the cause of loss of load. Again, the ranking might be predictable from the topological perspective but when we extend it to half load conditions, the results change as shown in Fig. 9. This is because during different load conditions, the importance of components is different. Although, it includes the same component, the ranking is different and it verifies the accuracy and effectiveness of Bayesian approach.

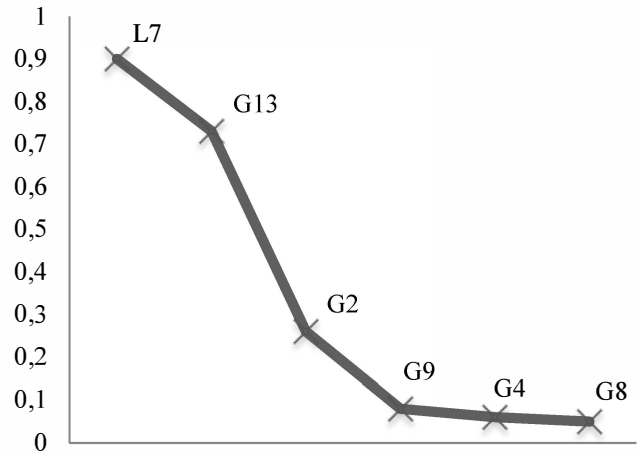

Fig. 8. Component ranking for bus 13 at full load condition

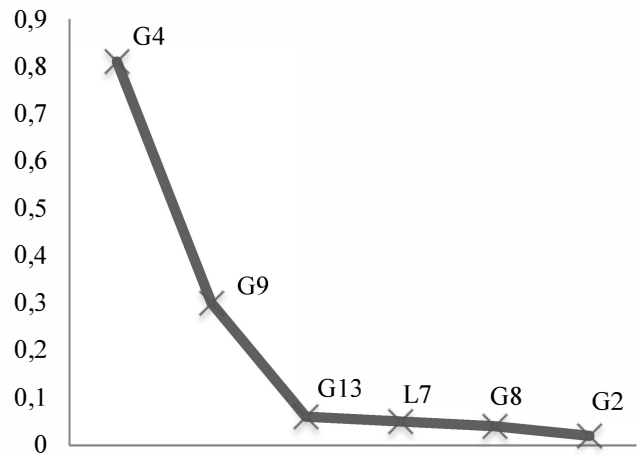

Fig. 9. Component ranking for bus 13 at half load condition

Ranking of load buses of RTS was performed in a similar manner as for RBTS. The RTS consists of 17 load buses. The complexity with first approach increases with increase in number of load buses. So, the ranking is performed using the second approach and is illustrated in Table II.

TABLE II RANKING OF LOAD BUSES OF IEEE RTS-24

| Bus | 13 | 20 | 18 | 15 | 7 | 2 | 1 | Rest |
|---|---|---|---|---|---|---|---|---|
| $(P(B=F|LOL=T))$ | 0.66 | 0.54 | 0.33 | 0.25 | 0.20 | 0.19 | 0.12 | < 0.1 |

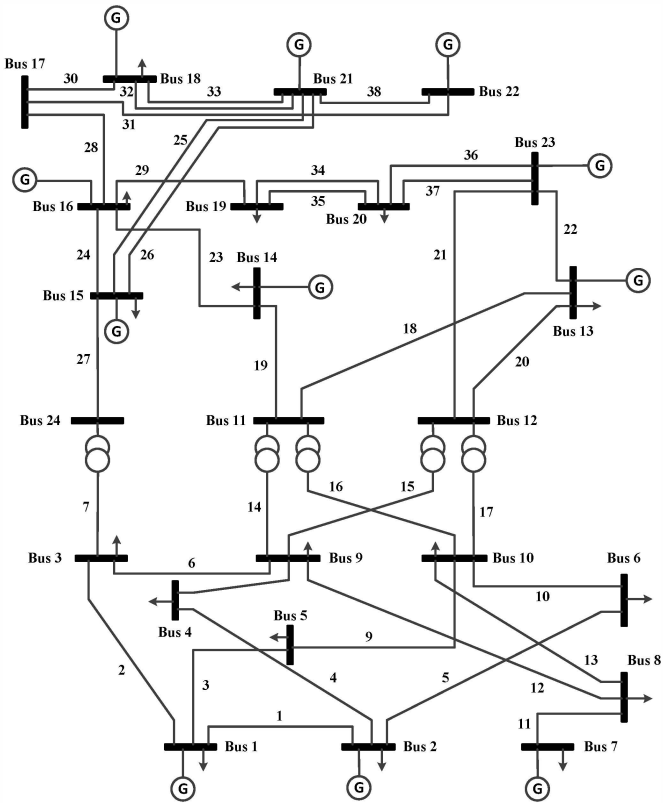

Fig. 6. Second test case: Single line diagram of IEEE RTS-24

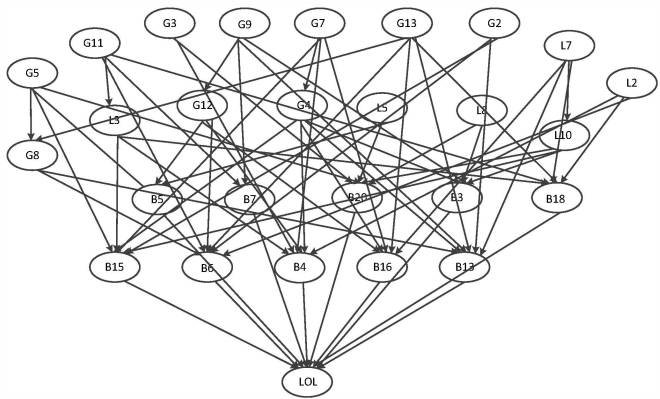

Fig. 7. Bayesian structure of IEEE RTS-24 bus system

## V. CONCLUSION

A detailed reliability analysis is made using mutual information based Bayesian approach. The robustness of desired methodology was proved by considering RBTS as well as IEEE RTS-24 bus system. The Bayesian network was constructed using rare outage events, first by a general structure, and then using mutual information for better study. With this approach, it makes it possible to utilize expert decision on the relationships between events and to take their uncertainty into account. The use of importance sampling in this study resulted in identifying system components that have low probabilities. This resulted in constructing a meaningful Bayesian network for reliability analysis. As a part of future study, the approach is to be extended to a real-time system.